\documentclass[
               aps,prd,
               showpacs,preprintnumbers,amsmath,amssymb,
               nofootinbib]{revtex4}
\usepackage{bm}
\def\be{\begin{equation}}
\def\ee{\end{equation}}
\def\bea{\begin{eqnarray}}
\def\eea{\end{eqnarray}}
\def\p{\partial}

\def\3g{{{}^{(3)}g}}
\def\A{{\cal{N}}}
\def\h{{\cal{H}}}
\def\z{{\psi}}

\renewcommand\({\left(}
\renewcommand\){\right)}
\renewcommand\[{\left[}
\renewcommand\]{\right]}

\newcommand\eq[1]{Eq.~(\ref{#1})}
\newcommand\eqs[2]{Eqs.~(\ref{#1}) and (\ref{#2})}

\newcommand\sub[1]{_{\rm #1}}

\begin{document}
%
%
\title{A general  proof of the  conservation of the 
curvature perturbation}
\author{David H.~Lyth}
\affiliation{Physics Department, University of Lancaster, Lancaster LA1 4YB, 
United
Kingdom}\author{Karim A.~Malik}
\affiliation{Physics Department, University of Lancaster, Lancaster LA1 4YB, 
United
Kingdom}
\author{Misao Sasaki}
\affiliation{Yukawa Institute for Theoretical Physics, Kyoto University, Kyoto 
606-8502, Japan}
\date{\today}
\begin{abstract}

Without invoking a perturbative expansion, we define the cosmological
curvature perturbation, and consider its behaviour assuming that the
universe is smooth over a sufficiently large comoving scale. The
equations are simple, resembling closely the first-order equations,
and they lead to results which generalise those already proven in
linear perturbation theory and (in part) in second-order perturbation
theory.  In particular, the curvature perturbation is conserved
provided that the pressure is a unique function of the energy density.

\end{abstract}
\pacs{98.80.-k \hfill JCAP05 (2005) 004, astro-ph/0411220v3}
\preprint{YITP-04-67}%
\maketitle 
%
\section{Introduction}

In this note we define and study a non-linear generalisation of the
linear cosmological `curvature perturbation' which is commonly denoted
by $\zeta$. Just as in linear theory, we find that the curvature
perturbation is conserved during any era when the locally-defined
pressure is a unique function of the locally-defined energy density,
but is otherwise time-dependent.

Before describing the calculation we mention the motivation which is
actually two-fold.  The first motivation concerns the comparison of
theory with observation. Observation has now established the existence
of a nearly Gaussian and scale-invariant spatial curvature
perturbation, present already before comoving cosmological scales come
inside the Hubble distance (enter the horizon).  The curvature
perturbation presumably is generated somehow by the perturbation of a
scalar field, that perturbation in turn being created on each scale at
horizon exit during inflation. Several mechanisms for achieving this
have been proposed.  In all of them it is supposed that the curvature
perturbation, defined on suitable spacetime slices, is conserved
(starting from the time when the creation mechanism ceases) until the
approach of horizon entry.  The conservation is supposed to hold by
virtue of the condition, either proven or just assumed, that the
pressure is a unique function of the energy density.  This is the
`adiabatic pressure' condition, that in first order perturbation
theory leads to $\delta P/\delta \rho = \dot P/\dot\rho$.

The conservation of the curvature perturbation for adiabatic pressure
has been demonstrated at linear order for an appropriate slicing of
spacetime. That slicing can be taken to be the either the `comoving'
one (the one orthogonal to comoving worldlines) or the uniform-density
one or the uniform-proper-expansion (uniform-Hubble) one; it does not matter
which because in linear theory the three are known to coincide in the
large-scale limit~\cite{Bardeen80,KodSas}.

These results are all that one needs at the moment, but with better
data in the future it will be necessary to go to second order. This is
because primordial non-Gaussianity, which may well be big enough to be
detected, is typically generated only at the second-order level.  The
study of second-order gauge invariant perturbations started only quite
recently~\cite{Mukhanov,Bruni} and has been hampered by it's sheer
complexity. The prospect of observing non-Gaussianity in the near
future has led to renewed activity in this
area~\cite{Maldacena,Acquaviva,Noh,MW,Nakamura} but the second-order
approach has not yet been carried through to the point where it
sufficiently generalises the first-order large-scale results.

As serious as its incompleteness is the fact that even on very large
scales the second order calculation is complicated, losing as a result
the physical transparency of the first-order calculation. This brings
us to our first motivation; we will provide a non-linear understanding
of the large-scale situation (applying therefore to all orders in the
perturbative expansion) which follows the same lines as the
first-order discussion and is not much more complicated. We will show
explicitly how to recover powerful second-order results from this
approach, without going through the complicated second-order
formalism.

Our second motivation is more philosophical, concerning the
possibility that inflation lasted for an exponentially large number of
$e$-folds. In that case the expanding Universe which we observe is
part of a huge region, which at the classical level was homogeneous
before it left the horizon.  When comparing the theory with
observation, very distant parts of this huge region are irrelevant,
because one can and should formulate the theory within a comoving box
whose present size is just a few powers of ten bigger than the Hubble
distance.  Linear theory with the small second-order correction is
then valid within the box, which is all that one needs for practical
purposes.  Still, it is of interest to understand the nature of the
universe in very distant regions.  In this connection it is important
to recognise that a Gaussian `perturbation' with a scale-invariant
spectrum actually becomes indefinitely large in an indefinitely large
region.  This means that one should try to understand the cosmological
perturbations without invoking the perturbative expansion, which is
precisely what we do here.

We work with the ADM formalism~\cite{ADM}. Historically, the
ADM-formalism was employed to study non-linearities already by Bardeen
in~\cite{Bardeen80}, albeit in the context of perturbation theory.
Non-perturbative studies of fluctuations followed in~\cite{SB} (see
also \cite{Afshordi,Shellard}).
In~\cite{ShAs} the relation between PPN-formalism and linear
perturbation theory was established using the ADM-formalism and
in~\cite{ShSa} the formation of primordial black holes was studied
numerically.

In addition, we should mention the existence of related work done by
the Russian school, which is based on construction of solutions of
Einstein equations having locally Friedmann-like behaviour near the
singularity, the so-called "quasi-isotropic" approach.  It was
pioneered by Lifshits and Khalatnikov in 1963~\cite{LifKha},
generalised to the de Sitter case~\cite{Staron} and has been
elaborated recently~\cite{KhaKam,KhaKamMarSta}.

This paper is organised as follows.  In Section~\ref{geometry} we
define the metric of our spacetime, specify the assumptions for its
validity, and describe the basic geometrical properties under our
assumptions.  In Section~\ref{sect_dyn} we show the conservation of a
non-linear generalisation of the curvature perturbation on very large
scales. In Section~\ref{cons_quant} we express our conserved quantity
in a slice-independent manner and relate it to the $\Delta N$ formula
derived in~\cite{SaSt95} (see also
\cite{Starobinsky1982,Starobinsky1986}). We conclude in the final
section. In Appendix~A we give the components of the relevant tensors
and in Appendix~B we give the spatial traceless components of the
Einstein equations.

\section{Geometry and energy conservation}
\label{geometry}
\subsection{Metric}
\label{metric_form}

We use the standard $(3+1)$-decomposition of the metric (ADM formalism), 
which applies to any smooth spacetime~\cite{ADM}:
\bea
ds^2 = -\A^2dt^2+ \gamma_{ij}(dx^i+\beta^idt)(dx^j+\beta^jdt)
\,,
\label{metric0}
\eea
where $\A$ is the lapse function, $\beta^i$ the shift vector, and
$\gamma_{ij}$ the spatial three metric. 
(Greek indices will take the values $\mu,\nu=0,1,2,3$, Latin indices
$i,j=1,2,3$. The spatial indices are to be raised or lowered by $\gamma^{ij}$
or $\gamma_{ij}$.) In this $(3+1)$-decomposition, the unit time-like vector
normal to the $x^0=t=$const. hypersurface $n^\mu$ has the components,
\begin{eqnarray}
n_\mu=\left[-\A,0\right],
\quad n^\mu=\left[\frac{1}{\A},-\frac{\beta^i}{\A}\right].
\end{eqnarray}

We write the 
3-metric, $\gamma_{ij}$, as a product of two terms,
\be
\label{defgammaij}
\gamma_{ij}\equiv e^{2\alpha}\tilde\gamma_{ij} \,,
\ee
where $\alpha$ and $\tilde\gamma_{ij}$ are functions of
the spacetime coordinates $(t,x^i)$, and $\det[\tilde\gamma_{ij}]=1$. 
Because of the latter  condition, the first factor is a locally-defined
scale factor which we denote by $\tilde a$,
\be
e^\alpha \equiv \tilde a
\,.
\label{locala}
\ee

We are interested though in the inhomogeneity of $\alpha$, 
and so we  factor out from $e^\alpha$ 
a global  scale factor $a(t)$ and a  perturbation $\psi$,
\be 
e^\alpha = a(t) e^{\psi(t,x^i) }
\,.
\label{alpha}
\ee
We  assume that  $\psi$ vanishes 
somewhere in the observable Universe (say at our location).
This makes $a(t)$ the scale factor for our part of the universe,
and ensures that $\psi$ is a small perturbation throughout the observable
Universe.
(If we are surrounded by a super-large region within which $\psi$ becomes
large, then the appropriate scale factor for a generic observer
should be $\tilde a$ evaluated at their own location, making again
$\psi$ small in their vicinity.)

In a similar way the matrix
$\tilde\gamma_{ij}$ can be factored,
\be
\tilde\gamma  \equiv I e^H
\,,
\ee
where $I$ is the unit matrix.
The condition $\det(\tilde\gamma)=1$  ensures that the matrix
$H$ is traceless, which follows from the relation
 $\det(\exp(M))=\exp({\rm{tr}}(M))$, valid for any symmetric
matrix  $M$.
In our part of the universe, with coordinates corresponding to the 
usual gauges, $H_{ij}$ is a small perturbation.

\subsection{Gradient expansion}
\label{gradexpand}

Cosmological perturbation theory expands the exact equations in powers
of the perturbations, keeping only terms of a finite order.  In
particular, first-order perturbation theory linearises the exact
equations.  We will not use this perturbative approach, but instead
will use the gradient expansion method~\cite{SB,Deruelle,ShSa}, which
is an expansion in the spatial gradient of these inhomogeneities.  To
be precise, we focus on some fixed time, and multiply each spatial
gradient $\partial_i$ by a fictitious parameter $\epsilon$, and expand
the exact equations as a power series in $\epsilon$. Then we keep only
the zero- and first-order terms and finally set $\epsilon=1$.

In the perturbative approach the fictitious
parameter $\epsilon$ would be multiplying  the perturbations.
Working to linear
order in the $\epsilon$ of the gradient expansion
obviously reproduces that 
subset of the linear perturbation theory expressions which can be derived
by considering only those equations which have at least one spatial gradient
acting on each perturbation. This is why our results will
closely resemble those of linear perturbation theory.

The gradient expansion is useful when every quantity can be
assumed to be smooth on some sufficiently large scale with coordinate size
$ k^{-1}$. If we are to model our actual universe by this smoothed universe,
it is necessary to implement a smoothing procedure at the level of the
field equations, either of Einstein gravity or of any alternative theory.
However, smoothing the smaller scale inhomogeneities is a delicate issue, 
which is beyond the scope of the present paper. Here, we simply assume that
there exists some kind of smoothing that can give a good approximation to
the actual universe on coordinate scales greater
than $k^{-1}$.\footnote{In linear perturbation theory 
it is appropriate to use a Fourier expansion and then smoothing corresponds
to dropping Fourier components with wavenumber bigger than $k$,
but we have no use here for the Fourier expansion. Still, it is useful to
keep the case of the Fourier expansion in mind.}  
Focusing on the observable Universe, this  corresponds to 
 a  comoving smoothing scale of  physical size $a(t)/k$. 
Then, instead of introducing
$\epsilon$ as a  formal multiplier of the spatial gradients, we can 
make the identification
\be
\epsilon \equiv k/aH
\,,
\ee
where $H=\dot a/a$ is the Hubble parameter.\footnote{If we are
surrounded by a super-large region, $a$ should be replaced by the
local scale factor $\tilde a(t,x^i)$, and $H$ by the local Hubble parameter
$\tilde H(t,x^i)$ that we define later.}
At a fixed time the limit $\epsilon\to 0$ corresponds to $k\to 0$.
In Hubble units, the typical value
of the gradient of a quantity $f$ will be $\epsilon f$. 

Our key physical assumption is that in the limit $\epsilon\to 0$,
corresponding to a sufficiently large smoothing scale, the universe
becomes {\em locally} homogeneous and isotropic (a FLRW universe). (By
`locally' we mean that a region significantly smaller than the
smoothing scale, but larger than the Hubble scale, is to be
considered.)  The Hubble distance is the only geometric scale in the
unperturbed universe.  In the perturbed Universe there is in addition
the scale $1/k$ under consideration, and possibly other scales
provided by the stress-energy tensor. Unless one of the latter scales
is bigger than $1/k$, local homogeneity and isotropy will be a good
approximation achieved throughout the entire super-horizon era $k\ll
aH$, and the results of this paper will be valid throughout that era.
Because cosmological scales are so large, one expects this `separate
universe' hypothesis \cite{SaTa,WMLL,LW} to be valid for them, ensuring the
maximum regime of applicability for our results.

An immediate consequence of our assumption is that the {\em locally
measurable} parts of the metric should reduce to those of the
FLRW. Thus there exists an appropriate set of coordinates
with which the metric of any local region can be written
\be
ds^2 = -dt^2 +  a^2(t) \delta_{ij} dx^i dx^j
\,.
\ee
(We took this metric to be spatially flat, which is the expectation
from inflation and agrees with observation; a small homogeneous curvature
would make no difference.) 

Let us see what this implies for the metric components. 
In the limit $\epsilon\to0$, the above local metric should be
globally valid. This implies that the metric component $\beta_i$
 vanishes in this limit,
 $\beta_i=O(\epsilon)$.\footnote{We adopt the traditional
mathematics notation \cite{hardy}, according
to which $f=O(\epsilon^n)$ means that $f$ falls like $\epsilon^n$ {\em
or faster}.} It may be noted, however, that this is not really a
necessary condition but rather a matter of choice of coordinates
for convenience.

What about the quantity $\tilde \gamma_{ij}$?  A homogeneous
time-independent $\tilde\gamma_{ij}$ can be locally transformed away
by choice of the spatial coordinates, but a homogeneous time-dependent
$\tilde\gamma_{ij}$ is forbidden because it would not correspond to a
FLRW universe. We therefore require $\dot{\tilde\gamma}=O(\epsilon)$.
In Appendix~\ref{app_2} though, we show that $\dot{\tilde\gamma}$ decays
 like $\tilde a^{-3}$ in Einstein gravity if it is really
linear in $\epsilon$.  Taking the usual view that decaying
perturbations are to be ignored, we conclude that $\dot{\tilde\gamma}$
will be of second order in $\epsilon$. 

The conditions on the metric components are therefore
\bea 
\beta_i 
&=& O(\epsilon)\,,
 \label{betacon}\\ 
\dot{\tilde\gamma}_{ij} &=&
 O(\epsilon^2)
\,.  
 \label{gammacon} 
\eea 
There is no requirement on
 $\psi$ and $\A$ since they are not locally observable.
We note that in alternative theories of
gravity, the assumption $\dot{\tilde\gamma}=O(\epsilon^2)$
may not be as natural as in the Einstein case.
 Nevertheless, we assume this condition. In other words, we 
implicitly focus on a class of gravitational theories in which the
condition $\dot{\tilde\gamma}=O(\epsilon^2)$ is consistent with
the field equations.

In view of \eq{betacon}, the line element
simplifies, giving
\be
ds^2 = -  \A^2 dt^2 +2\beta_i dx^i dt + \gamma_{ij} dx^i dx^j
\,.
\label{metric1}
\ee

\subsection{{Energy conservation}}
\label{energy_cons}

By virtue of the separate universe assumption, the energy momentum
 tensor will have the perfect fluid form
\be
\label{defTmunu}
T_{\mu\nu}\equiv \left(\rho+P\right)u_\mu u_\nu
+g_{\mu\nu}P \,,
\ee
where $\rho=\rho(x^\mu)$ is the energy density and $P=P(x^\mu)$ is the
pressure. 

First let us choose the spatial coordinates that comove with the fluid. 
That is, the threading of the spatial coordinates such that
the threads $x^i=$constant coincide with
the integral curves of the 4-velocity $u^\mu$
(the comoving worldlines). Hence,
\begin{eqnarray}
v^i=\frac{u^i}{u^0}\left(=\frac{dx^i}{dt}\right)=0\,.
\label{comove}
\end{eqnarray}
The components of the 4-velocity in these coordinates
are 
\begin{eqnarray}
&&u^\mu = \left[\frac{1}{\sqrt{\A^2-\beta^k\beta_k}}, 0\right]
=\left[\frac{1}{\A},0\right]+O(\epsilon^2)\,,
\nonumber\\
&&u_\mu = \left[-\sqrt{\A^2-\beta^k\beta_k}, 
\frac{\beta_i}{\sqrt{\A^2-\beta^k\beta_k}}\right]
=\left[-\A,\frac{\beta_i}{\A}\right]+O(\epsilon^2)\,.
\end{eqnarray}
The expansion of $u^\mu$ in the comoving coordinates, $v^i=0$,
is given by
\begin{eqnarray}
\theta\equiv\nabla_\mu u^\mu=
\frac{1}{\sqrt{-g}}\partial_\mu\left(\sqrt{-g}u^\mu\right)
=\frac{1}{{\cal N}{e^{3\alpha}}}\partial_0
\left({\cal N}e^{3\alpha}u^0\right)
=\frac{1}{{\cal N}{e^{3\alpha}}}\partial_t
\left(\frac{{\cal N}e^{3\alpha}}{\sqrt{{\cal N}^2-\beta^i\beta_i}}\right).
\end{eqnarray}
Note that $\tilde\gamma_{ij}$ does not appear in the above
expression because $\det\tilde\gamma_{ij}=1$.
The relation between the coordinate time $x^0=t$ and the proper time
$\tau$ along $u^\mu$ is
\begin{eqnarray}
\frac{dt}{d\tau}=u^0=\frac{1}{\sqrt{{\cal N}^2-\beta^i\beta_i}}\,.
\end{eqnarray}
The energy conservation equation, 
\begin{eqnarray}
-u_\mu \nabla_\nu T^{\mu\nu}
=\left[\frac{d}{d\tau}\rho+\left(\rho+P\right)\theta\right]=0\,,
\label{econs}
\end{eqnarray}
reduces therefore to
\be
\sqrt{\A^2-\beta^k\beta_k}
\left[\frac{d}{d\tau}\rho+\left(\rho+P\right)\theta\right]
=\dot\rho+3\left(\rho+P\right)\dot\alpha
+O(\epsilon^2)=0\,,
\label{cont}
\ee
where 
\be
\label{exp}
\theta=\frac{3\dot\alpha}{\A}+O(\epsilon^2)\,.
\ee
It is important to note that the expansion of the hypersurface normal
$n^\mu$ is given by
\begin{eqnarray}
\theta_n\equiv\nabla_\mu n^\mu
=\frac{3\dot\alpha}{\A}
-\frac{1}{\A\,e^{3\alpha}}\partial_i\left(e^{3\alpha}\beta^i\right)\,.
\label{nexp}
\end{eqnarray}
Thus $\theta$ and $\theta_n$ are equal to each other at linear order
in $\epsilon$. Note that the above argument uses only the energy
conservation law, hence applies to any gravitational theory as long as
the energy conservation law holds.

Here let us point out a couple of immediate but important
implications of the above. The first point is that the equivalence of
$\theta$ and $\theta_n$ for the comoving threading readily implies the
equivalence of $\theta$ and $\theta_n$ for any choice of threading for
which $\beta^i=O(\epsilon)$. This is because the change of the
threading affects the numerical value of $\theta_n$ at a given world
point $(t,x^i)$ only by terms of $O(\epsilon^2)$ as is clear from
Eq.~(\ref{nexp}).  The second point is that, the above argument
applies not only to the total fluid but also to any sub-component of
the fluid, provided that it does not exchange energy with the rest,
and that the comoving threading with respect to that component
satisfies the condition $\beta^i=O(\epsilon)$, that is, if the
3-velocity $v^i$ remains $O(\epsilon)$ for any threading with
$\beta^i=O(\epsilon)$.

Once we have the equivalence of $\theta$ and $\theta_n$,
it is convenient to introduce the notion of a local `Hubble parameter'
 $3\tilde H \equiv \theta_n$,
\begin{eqnarray}
\tilde H=\frac{1}{3}\theta_n 
&=& \frac 1\A \( \frac{\dot a}{a} + \dot\psi \) +O(\epsilon^2)\,.
\label{localH}
\end{eqnarray}
As is shown in Appendix~\ref{energy_momentum_app}, we then
recover a local Friedmann equation once
we appeal to the Einstein equations.

\section{The evolution of the curvature perturbation}
\label{sect_dyn}

Now let us investigate the evolution of the curvature perturbation
$\psi$. So far, we have not specified the choice of the time-slicing.
Below we shall consider some typical choices of the time-slicing
separately.

\subsection{Uniform-density slicing}

In this subsection we consider the uniform-density slicing, denoting 
$\psi$ on this slicing by $-\zeta$. We shall need only the condition
$\beta=O(\epsilon)$, not the other condition 
$\dot{\tilde\gamma} = O(\epsilon^2)$.

Following the linear treatment of \cite{WMLL}, we avoid in this
subsection the assumption of Einstein gravity.  Instead we just
consider some energy-momentum tensor $T_{\mu\nu}$, which satisfies
$\nabla_\nu T^{\mu\nu}=0$ corresponding to energy-momentum
conservation.  This is useful in two ways. First, it allows us to deal
if desired with just one component of the cosmic fluid instead of the
total.  Second, our results will apply to the case (arising for
instance in RSII cosmology \cite{RanSu}) that Einstein gravity is
actually modified.

Throughout this paper we are working to first order in $\epsilon$.
We  take the anisotropic stress of the fluid to be negligible at that order.
In other words, the anisotropic stress is supposed to be of second order
in $\epsilon$. (This can be verified in specific cases, in particular if
the fluid consists of a gas and/or scalar fields.)
Then, because there is a unique 
local expansion rate $\tilde H$ to linear order in $\epsilon$, 
the local energy conservation equation (\ref{cont})
has the unperturbed form to this order,
\be
\frac{d}{d\tau}\rho= -3 \tilde H \left(\rho+P\right) + O(\epsilon^2)
\,.
\label{eecons}
\ee
Multiplying each side by $\A$ this becomes
\be
\label{conti}
\frac{\dot a}{a} + \dot\psi 
= -\frac13 \frac{\dot\rho}{\rho + P} + O(\epsilon^2)
\,.
\ee

At each point this equation is valid independently of the slicing.
Now let us go to 
the uniform-density time-slicing, and denote $\psi$ on this slicing
by $-\zeta$. If, to first order in $\epsilon$,
 $P$ is a unique function of $\rho$ (the `adiabatic pressure' condition),
then Eq.~(\ref{conti}) shows that
$\dot\psi$ is spatially homogeneous to first order.
Since $\psi$ is supposed to vanish at say our position, 
though the position can be chosen arbitrarily,
this means that $\psi$ on
uniform-density slices is time-independent to first order,
\begin{eqnarray}
\label{dotz}
-\dot\psi=\dot\zeta = O(\epsilon^2)\,.
\end{eqnarray}
At this stage, as noted before, $\zeta$ can refer to the total cosmic
fluid, or to a single component which does not exchange energy with
the remainder.

\subsection{The comoving and uniform-Hubble slicings}
\label{sect_gauges}

Now we invoke Einstein gravity, and show that the comoving and
uniform-Hubble slicings coincide to first order in $\epsilon$ with the
uniform-density slicing. By `comoving slicing' we mean the one
orthogonal to the comoving worldlines (it should be perhaps called the
velocity-orthogonal slicing from the general relativity point of view,
but the terminology `comoving slicing' is usual in cosmology).  As we
have already decided to use the comoving worldlines as the threading
this fixes the gauge completely, and we call it the comoving gauge.

By invoking the comoving slicing we are setting the vorticity of the
fluid flow equal to zero, since the slicing exists if and only if that
is the case.  This is reasonable because vorticity is not generated
from the vacuum fluctuation during inflation.  Also, for a perfect
fluid, there is a vorticity conservation law in arbitrary
spacetime~\cite{HE}, which states that the magnitude of the vorticity
vector is inversely proportional to $S\exp[\int dP/(\rho+P)]$ along
each fluid line, where $S$ is the cross-sectional area of a congruence
of the fluid orthogonal to the vorticity vector.  Thus, the vorticity
would become negligible a few Hubble times after horizon exit on each
scale, even if it were somehow generated during inflation.

The Einstein equations are
\be 
G_{\mu\nu} = 8\pi G T_{\mu\nu} \,, 
\ee
where $T_{\mu\nu}$ refers now to the total fluid, and $G_{\mu\nu}$ and
$G$ are the Einstein tensor and Newton's constant, respectively.  We
consider its components in Appendix A3.  The $(0,i)$-component gives
in the comoving gauge
\be
\label{G0i}
\partial_i \tilde H = O(\epsilon^3)
\,,
\ee
and the  
 $(0,0)$-component gives the local Friedmann equation
\be 
\tilde H^2 = \frac{8\pi G}{3} \rho + O(\epsilon^2) 
\,,
\ee
leading to $\partial_i \rho = O(\epsilon^3)$. This makes 
 the typical magnitudes
of $\delta \rho$ and $\delta \tilde H$ go
 like $\epsilon^2$. In other words, the uniform-density, uniform Hubble
and comoving slices coincide to linear order in $\epsilon$.
These results are the same as those of 
 linear perturbation theory.

Knowing that $\tilde H$ and $\rho$ are both spatially homogeneous, we
can learn about the lapse function by writing the energy conservation
equation \eq{eecons} in terms of coordinate time,
\be
\frac1\A \dot\rho= -3 \tilde H \left(\rho+P\right) + O(\epsilon^2)
\,.
\ee
Since $\dot\rho$ and $\tilde H$ are spatially uniform we learn that
$\A$ is of the form
\be
\A =   \frac{A(t)}{\rho(t)+ P(t,x^i) } + O(\epsilon^2)
\,,
\label{calneq}
\ee
where $A(t)$ can be any function which makes $\A$ positive definite.
If the pressure is adiabatic, $\A$ is independent of position and we
can choose $\A=1$. In that case, the comoving gauge is also a
synchronous gauge to first order in $\epsilon$.

If the pressure is not adiabatic we can write on the uniform energy density
slicing
\be
\A = \frac{\rho(t) + P(t)}{\rho(t) + P(t,x^i) }
\,.
\ee
This is the non-linear generalisation of the 
 known result in first-order perturbation theory,
\be
\A = 1 - \frac{\delta P}{\rho + P}
\,,
\ee
where $\delta P$ is the pressure perturbation on uniform-density 
slices (the `non-adiabatic' pressure perturbation).\footnote{As we
 were preparing the present paper, a related one appeared~\cite{kmnr}
 claiming that the spatial variation of the coordinate expansion rate
in the non-adiabatic case
could change the presently-accepted predictions for observable quantities.
We disagree with this
conclusion, since that variation is already implicitly present in any correct
formulation of cosmological perturbation theory.}

As shown in Appendix~\ref{app_2},
the traceless part of the spatial components of the Einstein equations
is $O(\epsilon^2)$, while the trace part gives no additional 
information.

\section{Gauge transformations and ${\bf{\Delta N}}$ formula}
\label{cons_quant}

As we are working to first order in $\epsilon$, the threading is
unique in the sense that all the threadings are equivalent to
the comoving threading as discussed in Section~\ref{energy_cons},
leaving only the slicing to be fixed. In this section 
we consider the effect of a change of slicing on the curvature
perturbation. 

Let us define the number of $e$-foldings of expansion 
along an integral curve of the 4-velocity (a comoving worldline):
\begin{eqnarray}
N(t_2,t_1;x^i)\equiv\frac{1}{3}\int_{t_1}^{t_2}\theta\,\A dt
=-\frac{1}{3}\int_{t_1}^{t_2}dt
\left.\frac{\dot\rho}{\rho+P}\right|_{x^i}\,,
\label{Ndef}
\end{eqnarray}
where, for definiteness, we have chosen the spatial coordinates
$\{x^i\}$ to be comoving with the fluid. The essential point to be
kept in mind is that this definition is purely geometrical,
independent of the gravitational theory one has in mind, and applies
to any choice of time-slicing.

{}From \eq{localH} we find
\be
\psi(t_2,x^i)-\psi(t_1,x^i)
=N(t_2,t_1;x^i)-\ln\left[\frac{a(t_2)}{a(t_1)}\right]\,.
\label{evolve}
\ee
Thus we have the very general result that the change in $\psi$, going
from one slice to another, is equal to the difference between the
actual number of $e$-foldings and the background value
$N_0(t_2,t_1)\equiv\ln[a(t_2)/a(t_1)]$.  One immediate consequence of
this is that the number of $e$-foldings between two time slices will
be equal to the background value, if we choose the `flat slicing' on
which $\psi=0$. (This slicing is of course truly flat only if
$\tilde\gamma_{ij}=\delta_{ij}$.) Thus the flat slicing is one of the
{\it uniform integrated expansion} slicings~\cite{LW}.

Consider now two different time-slicings, say slicings $A$ and $B$,
which coincide at $t=t_1$ for a given spatial point $x^i$ of our
interest (i.e., the 3-surfaces $\Sigma_A(t_1)$ and $\Sigma_B(t_1)$ are
tangent to each other at $x^i$). Then the difference in the
time-slicing at some other time $t=t_2$ can be described by the
difference in the number of $e$-foldings. From Eq.~(\ref{evolve}), we
have
\begin{eqnarray}
\psi_A(t_2,x^i)-\psi_B(t_2,x^i)
&=&N_A(t_2,t_1;x^i)-N_B(t_2,t_1;x^i)
\nonumber\\
&\equiv&\Delta N_{AB}(t_2,x^i)\,,
\label{gaugetran}
\end{eqnarray}
where the indices $A$ and $B$ denote the slices $A$ and $B$,
respectively, on which the quantities are to be evaluated.  As
discussed in Section~\ref{applications}, this generalises, for large
scales only, the known result of the first- \cite{Bardeen80} and
second-order perturbation theory~\cite{Mukhanov,Bruni,MW}.

Now let us choose the slicing $A$ to be such that it starts on a flat
slice at $t=t_1$ and ends on a uniform-density slice at $t=t_2$, and
take $B$ to be the flat slicing all the time from $t=t_1$ to $t=t_2$.
Then applying Eq.~(\ref{gaugetran}) to this case, we have
\begin{eqnarray}
\psi_A(t_2,x^i)
&=&N_A(t_2,t_1;x^i)-N_0(t_2,t_1)=\Delta N_F(t_2,t_1;x^i)\,,
\end{eqnarray}
where $\Delta N_F(t_2,t_1;x^i)$ is the difference in the number of
$e$-foldings (from $t=t_1$ to $t=t_2$) between the uniform-density
slicing and the flat slicing.  This is a non-linear version of the
$\Delta N$ formula that generalises the first-order result of Sasaki
and Stewart~\cite{SaSt95}.

Now we specialise to the case $P=P(\rho)$. In this case,
Eq.~(\ref{evolve}) reduces to
\begin{eqnarray}
\psi(t_2,x^i)-\psi(t_1,x^i)
=-\ln\left[\frac{a(t_2)}{a(t_1)}\right]
-\frac{1}{3}\int^{\rho(t_2,x^i)}_{\rho(t_1,x^i)}\frac{d\rho}{\rho+P}\,.
\end{eqnarray}
Thus, there is a conserved quantity, which is independent of the
choice of time-slicing, given by
\begin{eqnarray}
-\zeta(x^i)\equiv\psi(t,x^i)
+\frac{1}{3}\int_{\rho(t)}^{\rho(t,x^i)}\frac{d\rho}{\rho+P}\,.
\label{nlconserve}
\end{eqnarray}
In the limit of linear theory, this reduces to the conserved
curvature perturbation in the uniform-density, uniform-Hubble, or 
the comoving slicing,
\begin{eqnarray}
-\zeta(x^i)={\cal R}_c(x^i)
=\psi(t,x^i)+\frac{\delta\rho(t,x^i)}{3(\rho+P)}\,.
\label{linear}
\end{eqnarray}

Finally, we mention that the generalisation of all the above results
to the case of arbitrary threading, not restricted by the condition
$\beta^i=O(\epsilon)$, is formally trivial.  Let us denote the general
spatial coordinates by $\{X^i\}$.  They are related to the comoving
coordinates $\{x^i\}$ by a set of coordinate transformations,
$x^i=F^i(t,X^i)$.  Then all the equations above are valid for an
arbitrary choice of threading by simply replacing the arguments $x^i$
of all the functions by $F^i(t,X^i)$.

\section{Applications to cosmological perturbation theory}
\label{applications}

In this section we make contact with cosmological perturbation theory,
showing how  our results both reproduce and extend
known second-order results.

Let us begin with the first-order case.
To first order in the perturbations the spatial metric becomes\footnote 
{The notation $\psi$ is that of \cite{MFB};
Kodama and Sasaki~\cite{KodSas} denote the same quantity by
 $\cal R$).}
\be 
g_{ij} = a^2(t) \[ \delta_{ij} \( 1+ 2\psi \) + H_{ij} \]
\,, 
\label{mw} 
\ee
In this context it is known (see for instance \cite{LW}) that (always
referring to the super-horizon regime) three slices coincide: uniform
density, uniform proper expansion (uniform Hubble, the expansion being
independent of the threading), and comoving slice. Also it is known
that the perturbation $\psi$ is independent of the threading. Finally,
defining the curvature perturbation $\zeta$ as the value of $-\psi$ on
this `triple coincidence' slicing, it is known that $\zeta$ is
conserved as long as pressure is a unique function of energy density.

Going to second order (with an appropriate definition of $\psi$) only
some of these statements have been verified. In particular, the
conservation of $\zeta$ was shown by Sasaki and Shibata \cite{ShSa} in
non-linear theory if $P/\rho$ is constant and by Malik and Wands
\cite{MW} in second order perturbation theory setting
$\tilde\gamma_{ij} = \delta_{ij}$ (i.e.\ ignoring the tensor), and by
Salopek and Bond during single-component inflation.  The main point of
our paper has been to show that all of them are in fact valid (with
again an appropriate definition of $\psi$).

Our finding is important  because it ensures that the
non-gaussianity of the curvature perturbation can be calculated once
and for all at the time of its creation, remaining thereafter constant
until horizon entry \cite{ournew}. 
This was the implicit assumption made by
Maldacena \cite{Maldacena} (see also \cite{SB2,MMOL,Shellard2,RS}); 
he calculated the non-gaussianity of the
curvature perturbation (to be precise, its bispectrum) a few Hubble
times after horizon exit in a single-field model, defining it on the
comoving slicing. 
It was also that of \cite{curv}, who calculated the
curvature perturbation just before curvaton decay, now on the uniform
density slicing, and that of \cite{kari} who calculated it at the end
of inflation in a two-component inflation model with a straight
inflaton trajectory (again on the uniform-density slicing).

The gauge transformations and gauge-invariant expressions that we derived
in the last section reproduce the  second-order results.
To see this,
 let us first consider two
definitions of the curvature perturbation in the literature.\footnote
{Another definition \cite{Acquaviva} is discussed elsewhere
\cite{ournew}.} 
In all cases we shall employ the notation that a generic
perturbation $g$ is split into first- and second-order parts according to
\be
g \equiv g_1 + \frac12 g_2
\,.
\ee
One definition, used by Maldacena (introduced by Salopek and Bond in
\cite{SB}) to calculate the non-Gaussianity generated by single-field
inflation, coincides with our definition of $\psi$,
\be 
e^{2\alpha} = a^2(t) e^{2\zeta} = a^2(t)\( 1+ 2\zeta + 2\zeta^2\)  \,.  
\ee
The other generalisation, employed by Malik and Wands (based on
\cite{Bardeen80,MFB,Bruni}), is different and we denote it by
$\zeta\sub{mw}$; it is
\be
e^{2\alpha}  = a^2(t) \( 1 + 2\zeta\sub{mw} \)
\,,
\ee
so that 
\be
\zeta\sub{mw} = \zeta + \zeta^2\,,
\ee
or equivalently
\be
\label{relation}
\zeta\sub{2mw} = \zeta_2 + 2 (\zeta_1) ^2 
\,.
\ee

Evaluated to first order in the perturbations, 
the gauge transformation \eq{gaugetran} reduces to the known  result
\cite{Bardeen80}
\be
\psi_A - \psi_B = H\Delta t
\,,
\ee
where $\Delta t$ is the time displacement between the slices.
If one of the slicings has uniform density and the other is flat
this gives a gauge-invariant definition  of $\zeta$,
\be
-\zeta = \psi +\frac{\mathcal{H}}{\rho_0'} \delta\rho
\,,
\ee
where a prime denotes differentiation with respect to conformal time and
$\mathcal{H}\equiv a'/a$, and the right hand side is evaluated on a generic
slicing.

To illustrate how things work at second order, we will just show how
\eq{gaugetran} reproduces the known second-order result
\cite{MW}.\footnote {That this happens is stated without proof in
\cite{kmnr}, and for the special case of adiabatic pressure in
\cite{LW}.}  Following along the lines of Lyth and Wands \cite{LW}, we
expand the integrated expansion to second order in a power series
expansion centred on the flat slicing of \eq{gaugetran},
\be
\label{Nexpand}
\delta N=\frac{\p N}{\p \rho}\delta\rho
+\frac{1}{2}\frac{\p^2 N}{\p \rho^2}\delta\rho^2\,.
\ee
Using \eqs{eecons}{Nexpand} this gives the perturbed expansion at
second order,
\be
\label{deltaN2}
\delta N_2=
\frac{\h}{\rho_0'}\delta\rho_2
-2\frac{\h}{{\rho_0'}^2}\delta\rho_1'\delta\rho_1
+\left(\h\frac{\rho_0''}{\rho_0'}-\h'\right)\left(\frac{\delta\rho_1}{\rho_0'}
\right)^2
\,,
\ee
where the right hand side is evaluated on flat slices.
The expression given in \eq{deltaN2} is then related to
$\zeta_{2\rm{mw}}$, if this is also evaluated on flat slices, by
$\zeta_{2\rm{mw}}=-\delta N_2+2\zeta_1^2$, which coincides with
\eq{relation} using \eq{nlconserve}, i.e.~$\zeta_2=-\delta N_2$.

\section{Conclusions}

Our central result is the existence of a conserved
\emph{non-perturbative} quantity $\zeta$, corresponding to the scalar
curvature perturbation in the perturbative case, which may be defined
on the uniform-density, uniform-expansion or comoving slices since
these coincide in the large scale limit. Locally, this statement
follows from the equations of the coordinate-free approach as given
for instance in \cite{cofree} and reviewed in \cite{LLBook}.\footnote{
While finishing version 2 of the present paper, a related
work on the evolution of perturbations appeared, using the covariant
approach \cite{Langlois2005}.}
We have here preferred to employ the usual coordinate approach.  This
is because the coordinate approach is used in practise in almost all
treatments of the evolution of perturbations during and after horizon
entry, owing to the relative ease with which it handles the effect of
particle collisions and free-streaming. Using the coordinate approach
has allowed us to make contact with existing second order perturbation
calculations, clarifying some previously mysterious connections
between those of different authors. Also, on very, very large scales,
it makes contact with the idea that the `perturbations' are actually
supposed to be random fields which can become arbitrarily large in an
arbitrarily large region, thus providing in some sense a
generalisation of the stochastic description of scalar field evolution
during inflation.  
Finally, the gradient expansion can be most easily implemented in the
coordinate approach, in particular if we want to go to the next order
in the gradient expansion.

\acknowledgments

KAM is grateful to David Wands and David Burton for useful
discussions. The work of MS is supported in part by Monbukagaku-sho
Grant-in-Aid for Scientific Research (S), No.~14102004.  The Lancaster
group is supported by PPARC grants PPA/G/O/2002/00469 and
PPA/V/S/2003/00104 and by EU grants HPRN-CT-2000-00152 and
MRTN-CT-2004-503369, and DHL is supported by PPARC grants
PPA/G/O/2002/00098 and PPA/S/2002/00272.

\appendix
\section{Tensor components}

\subsection*{The metric tensor}

The metric tensor is given by
\bea
\label{metric}
g_{00}&=&-\A^2+\beta^i\beta_i \,, \qquad g_{0i}=g_{i0}=\beta_i\,,\\
g_{ij}&\equiv&\gamma_{ij}=e^{2\alpha}\tilde\gamma_{ij}\,,
\eea
and 
\bea
g^{00}&=&-\frac{1}{\A^2} \,, 
\qquad g^{0i}=g^{i0}=\frac{\beta^i}{\A^2}\,,\\
g^{ij}&=&\gamma^{ij}-\frac{\beta^i\beta^j}{\A^2}=
\frac{1}{e^{2\alpha}}\tilde\gamma^{ij}-\frac{\beta^i\beta^j}{\A^2}\,,
\eea
where $\det[\tilde\gamma_{ij}]=1$ and $\beta^i=\gamma^{ij}\beta_j$.

\subsection*{3-geometry}

The unit timelike vector normal to the hypersurface $t=$const. 
is given by
\be
\label{defnormal}
n_\mu=\left[-\A,0,0,0\right]\,, \qquad
n^\mu=\left[\frac{1}{\A},-\frac{\beta^i}{\A}\right]\,.
\ee
This gives for the extrinsic curvature tensor \cite{MTW}
\bea
\label{excur}
K_{ij}&=&-n_{i;j}\\
&=&\frac{1}{2\A}\left[-\frac{\p}{\p t}\gamma_{ij}+\beta_{i|j}+\beta_{j|i}
\right]
\,,
\eea
which can be expressed as
\begin{eqnarray}
K_{ij}=-\frac{\theta_n}{3}\,\gamma_{ij}+A_{ij}\,,
\label{Kij}
\end{eqnarray}
where $\gamma^{ij}A_{ij}=0$ and $\theta_n$ is the
expansion of $n^\mu$,
\be
\theta_n=\nabla_\mu n^\mu\,.
\ee
On large scales we have $A_{ij}=O(\epsilon
)$
by virtue of our assumptions, and Eq.~(\ref{Kij}) reduces to
\be
K^i{}_{j}=-\frac{\theta_n}{3}\delta^i_j+O(\epsilon^2)
=\frac{\dot\alpha}{\A}\delta^{i}_{j}
+O(\epsilon^2)
=-\frac{1}{\A}\left(
\frac{\dot a}{a}+\dot\z\right)\delta^{i}_{j}
+O(\epsilon^2)\,.
\ee

In the case of a conformally flat 3-geometry,
i.e., for $\tilde\gamma_{ij}=\delta_{ij}$,
the intrinsic curvature on spatial 3-hypersurfaces is
expressed 
\be
{}^{(3)}R=-\frac{2}{e^{2\alpha}}
\delta^{ij}\left(\psi_{,i}\psi_{,j}+2\psi_{,ij}\right)\,.
\ee

\subsection*{Einstein tensor}
\label{Einstein}

Here we give the components of the Einstein tensor to first order in
$\epsilon$,  valid in an arbitrary gauge 
provided that the metric satisfies the conditions (\ref{betacon})
and (\ref{gammacon}). 

First let us consider the case of $\beta_i=0$. 
Then, the time derivative 
$\partial/\partial t$ is along the normal vector $n_\mu=(-{\cal N},0)$.
Then, for the $(0,0)$ and $(0,i)$ components of the Einstein tensor, 
we can apply the standard $(3+1)$-decomposition of the
Ricci curvature tensor (i.e.~the Gauss-Codacci equations),
which are essentially the Hamiltonian and momentum constraint equations.
They are given, for example, by Eqs.~(2.20) and (2.21) of \cite{ShSa}.
In our notation, they are
\begin{eqnarray}
G^0{}_0&=&-\frac{1}{2}\left(
{}^{(3)}R +\frac{2}{3}K^2-A^{ij}A_{ij}
\right),
\\
G^0{}_j&=&D_iA^i{}_j-\frac{2}{3}D_jK\,,
\end{eqnarray}
where $D_i$ is the covariant derivative with respect to the metric
$\gamma_{ij}$, and the extrinsic curvature $K_{ij}$ is given by 
Eq.~(\ref{Kij}) with $\theta_n=-K$.

Since $A_{ij}=O(\epsilon^2)$,
and ${}^{(3)}R$ involves at least  second derivatives of the 
metric tensor, these expressions reduce to
\begin{eqnarray}
G^0{}_0&=&-\frac{1}{3}K^2 + O(\epsilon^2),
\\
G^0{}_j&=&-\frac{1}{\cal N}\,\frac{2}{3}D_jK + O(\epsilon^3) \,,
\end{eqnarray}
where 
\begin{eqnarray}
K=-\frac{3}{\cal N}\left(\frac{\dot a}{a}+\dot \psi\right)+O(\epsilon^2)\,.
\label{traceK}
\end{eqnarray}

For the $(i,j)$ components, they can be decomposed into the trace and
traceless part.  Evaluation of the traceless part is slightly
involved.  We write down the corresponding components of the Einstein
equations in Appendix~\ref{app_2}, in which it is shown under our
assumptions that they are $O(\epsilon^2)$ and hence can be neglected.
The trace part reduces to
\begin{eqnarray}
\label{Gii}
G^i{}_i=\frac{2}{\cal N}\partial_tK-K^2\,.
\end{eqnarray}
Therefore
\begin{eqnarray}
G^i{}_j=\frac{1}{3}\left(\frac{2}{\cal N}\partial_tK-K^2\right)\delta^i_j
+O(\epsilon^2)\,.
\end{eqnarray}

It is easy to see that inclusion of $\beta_i=O(\epsilon)$ does not
change these results at all.  The above results can be regarded as the
$(3+1)$-decomposition with the hypersurface normal vector
$n^\nu=(1/{\cal N},0)$. Thus, for a general choice of the spatial
coordinates, we just have to perform the following replacements:
\begin{eqnarray}
G^0{}_0&\to& -n_\mu G^\mu{}_\nu n^\nu
=G^0{}_0+G^0{}_i\beta^i\,,
\\
{\cal N}G^0{}_j&\to&-n_\mu G^\mu{}_\nu h^{\nu}{}_j
={\cal N}G^0{}_j\,,
\\
G^{i}{}_j&\to& h^i{}_\mu G^\mu{}_\nu h^\nu{}_j
=G^0{}_j\beta^i+G^i{}_j\,.
\end{eqnarray}
So, the difference is of the form, $G^0{}_j\beta^i$, which is
$O(\epsilon^2)$.

To summarise, the components of the Einstein tensor
on large scales are given by
\begin{eqnarray}
G^0{}_0&=&-\frac{1}{3}K^2+O(\epsilon^2),
\label{G00app}
\\
G^0{}_j&=&-\frac{1}{\cal N}\,\frac{2}{3}D_jK+O(\epsilon^2)\,,
\label{G0iapp}
\\
G^i{}_j&=&\frac{1}{3}\left(\frac{2}{\cal N}\partial_tK-K^2\right)\delta^i_j
+O(\epsilon^2)\,,
\label{Gijapp}
\end{eqnarray}
where $K$ is given by Eq.~(\ref{traceK}).

Although not used in the present paper, it is useful to know
the form of $O(\epsilon^2)$ corrections. The correction terms
take a complicated form in general, but for the
spatially conformally flat metric ($\tilde\gamma_{ij}=\delta_{ij}$)
 with $\beta^i=0$, they take a relatively simple form.
The components of the Einstein tensor in this case,
valid to the accuracy of $O(\epsilon^2)$, are given by
\bea
G^0{}_{0}&=&-\frac{1}{3}K^2
+\gamma^{jk}\left(\z_{,j}\z_{,k}+2\z_{,jk}\right)
\,, 
\nonumber\\
G^0{}_{j}&=&-\frac{1}{\cal N}\,\frac{2}{3}K_{,j}
\,,
\nonumber\\
G^i{}_{j}&=&\frac{1}{3}\left(\frac{2}{\cal N}\partial_tK-K^2\right)\delta^i_j
\nonumber \\ 
&&\qquad+\gamma^{ik}
\left[\frac{1}{\A}\left(\A_{,k}\z_{,j}+\z_{,k}\A_{,j}
-\A_{,kj}\right)
+\z_{,k}\z_{,j}-\z_{,kj}\right]
 \,.
\eea
%

\subsection*{Energy-momentum tensor}
\label{energy_momentum_app}

The four-velocity is given by
\bea 
u^0 &=&
\left[\A^2-(\beta_k+v_k)(\beta^k+v^k)
\right]^{-1/2} \,, \\
u^i &=& u^0 v^i\,,
\eea
where $v^i$ is the spatial velocity, and 
\bea
u_0 &=&-u^0\left[
\A^2-\beta^k\left(\beta_k+v_k\right)
\right]\,, \\
u_i &=& u^0\left(v_i+\beta_i\right)\,,
\eea
and $v_i=\gamma_{ij}v^j$.
The components of the energy momentum tensor are
then given by
\bea
\label{T00}
T^0{}_{0}&=& -\left({u^0}\right)^2\left(\rho+P\right)
\left[\A^2-
\beta^k\left( v_k+\beta_k\right)\right]+P\,, \\
\label{T0i}
T^0{}_{i}
&=&\left({u^0}\right)^2\left(\rho+P\right)\left(v_i+\beta_i\right)\,, \\
\label{Tij}
T^i{}_{j}&=&\left({u^0}\right)^2
\left(\rho+P\right)v^i\left(v_j+\beta_j\right)
+\delta^i_{~j}\,P\,.
\eea

We note that $T^0{}_0=-\rho+O(\epsilon^2)$ if $\beta^i$ and $v^i$ are
both of $O(\epsilon)$. If we choose our spatial coordinates to be
comoving with the fluid, we have $v^i=0$.  
Then the $(0,i)$-component
of the Einstein equations tells us that $T^0{}_i=O(\epsilon)$, which
implies $\beta^i=O(\epsilon)$. 
 Hence the $(0,0)$-component of the
Einstein equations gives a local Friedmann equation at each 
spatial point $x^i$. 
In other words, as long as we are concerned with
Einstein gravity, it is unnecessary to assume $\beta^i=O(\epsilon)$,
but the only condition we need to obtain the local Friedmann equation
is the comoving condition $v^i=0$ for the spatial coordinates.

Defining the projection tensor 
\be
h^{\mu\nu}\equiv g^{\mu\nu} +u^\mu u^\nu\,,
\ee
i.e., projecting orthogonally to the velocity $u^\mu$,
we can write down the momentum conservation equation
\be
h_{\lambda\nu}\nabla_\mu T^{\mu\nu}=0\,,
\ee
which gives in components in the comoving gauge ($\beta^i=v^i=0$)
\be
\left(\rho+P\right)D_i\ln\A+D_iP=0\,,
\label{Neq}
\ee
which reduces to the linear result in the comoving gauge on large
scales. It may worth noting that this holds for general 
$\tilde\gamma_{ij}$.
 
The Raychauduri equation \cite{Raychaudhuri} in the comoving gauge
($\beta^i=v^i=0$) and assuming zero vorticity is given by 
\be
\frac{1}{\A}\dot\theta+\frac{1}{3}\theta^2
+\frac{1}{2}\left(\rho+3P\right)+O(\epsilon^2)=0\,.
\ee
With the identification of $\theta=3\tilde H$, this is equal to
the time derivative of the local Friedmann equation,
which can be also obtained by combining the $(0,0)$-component of 
the Einstein equations as given by \eqs{G00app}{T00},
 and the trace of the $(i,j)$-component
as given by \eqs{Gii}{Tij}.

\section{Evolution equations for ${\bf{\tilde\gamma_{ij}}}$}
\label{app_2}

Here we write down the traceless part of the spatial components of
the Einstein equations, that is, the evolution equations for
$\tilde\gamma_{ij}$.
They can be found, for example, in Eqs.~(2.11) and (2.12) of Shibata and
Sasaki~(SS)~\cite{ShSa}.
Their $\tilde A_{ij}$ is related to our $A_{ij}$
by $\tilde A_{ij}=e^{-2\alpha}A_{ij}$, or
$\tilde A^i{}_j\equiv\tilde\gamma^{ij}\tilde A_{ij}=A^i{}_j$.
Equation~(2.11) of SS is
\begin{eqnarray}
\partial_t\tilde\gamma_{ij}=-2{\cal N}\tilde A_{ij}
+\pounds_\beta\tilde\gamma_{ij}
-\frac{2}{3}\tilde\gamma_{ij}\partial_k\beta^k\,,
\end{eqnarray}
where $\pounds_\beta$ is the Lie derivative along $\beta^k$,
given for a second-rank tensor $Q_{ij}$ by
\begin{eqnarray}
\pounds_\beta Q_{ij}=\beta^k\partial_kQ_{ij}+Q_{ik}\partial_j\beta^k
+Q_{kj}\partial_i\beta^k\,.
\end{eqnarray}
Thus, with the assumption that $\beta^k=O(\epsilon)$,
the assumption $\partial_t\tilde\gamma_{ij}=O(\epsilon)$
is equivalent to $\tilde A_{ij}=O(\epsilon)$.

Equation~(2.12) of SS is
\begin{eqnarray}
\partial_t\tilde A_{ij}
&=&{\cal N}\left(K\tilde A_{ij}-2\tilde A_{ik}\tilde A^{k}{}_j\right)
+\frac{1}{e^{2\alpha}}
\left[{\cal N}\left({}^{(3)}R_{ij}-\frac{\gamma_{ij}}{3}{}^{(3)}R\right)
-\left(D_iD_j{\cal N}-\frac{\gamma_{ij}}{3}D^kD_k{\cal N}\right)
\right]
\nonumber\\
&&+\pounds_\beta\tilde A_{ij}-\frac{2}{3}\tilde A_{ij}\partial_k\beta^k
-8\pi G\frac{{\cal N}}{e^{2\alpha}}
\left(S_{ij}-\frac{\gamma_{ij}}{3}S^k{}_k\right)\,,
\label{dttilA}
\end{eqnarray}
where $S_{ij}$ is the spatial projection of the energy-momentum tensor,
\begin{eqnarray}
S_{ij}=T_{ij}\,,\quad S^k{}_k=\gamma^{k\ell}S_{k\ell}\,.
\end{eqnarray}
For a perfect fluid, or in the absence of anisotropic stress,
we have
\begin{eqnarray}
S_{ij}=(\rho+P)u_iu_j+\gamma_{ij}P\,.
\end{eqnarray}
Hence
\begin{eqnarray}
S_{ij}-\frac{\gamma_{ij}}{3}S^k{}_k=(\rho+P)(u^0)^2
\left[(v_i+\beta_i)(v_j+\beta_j)-\frac{\gamma_{ij}}{3}
(v_k+\beta_k)(v^k+\beta^k)\right],
\end{eqnarray}
which is of second order in $\epsilon$.

Assuming the anisotropic stress is negligible, then
Eq.~(\ref{dttilA}) reduces to
\begin{eqnarray}
\partial_t\tilde A_{ij}={\cal N}K\tilde A_{ij}+O(\epsilon^2)
=-3\,\partial_t\alpha\,\tilde A_{ij}+O(\epsilon^2)\,.
\label{tilA}
\end{eqnarray}
Therefore, if $\tilde A_{ij}=O(\epsilon)$,
Eq.~(\ref{tilA}) has a decaying solution,
\begin{eqnarray}
\tilde A_{ij}=e^{-3\alpha}C_{ij}\,,
\label{decaysol}
\end{eqnarray}
where $C_{ij}=O(\epsilon)$ and $\partial_t C_{ij}=O(\epsilon^2)$.
Assuming that this decaying solution is absent (or ignorable),
we have $\partial_t{\tilde\gamma}_{ij}=O(\epsilon^2)$, which we assumed
in the text.



\end{document}